# Spin-mediated hysteretic switching of unidirectional charge density waves by rotating magnetic fields

Zichao Chen[1,2#], Shiyu Zhu[1,2#]✉, Kailin Xu[1,2#], Ruwen Wang[1,2], Ningning Wang[1,2], Jianfeng Guo[1], Yunhao Wang[1,2], Xianghe Han[1], Zhongyi Cao[1,2], Jianping Sun[1,2], Hui Chen[1,2], Haitao Yang[1,2], Jinguang Cheng[1,2], Ziqiang Wang[3]✉, Hong-Jun Gao[1,2]✉

[1]Beijing National Laboratory for Condensed Matter Physics, Institute of Physics, Chinese Academy of Sciences, Beijing 100190, China

[2]School of Physical Sciences, University of Chinese Academy of Sciences, Beijing 100049, China

[3]Department of Physics, Boston College, Chestnut Hill, MA, 02467, USA

[#]These authors contributed equally: Zichao Chen, Shiyu Zhu and Kailin Xu.

✉ e-mail: syzhu@iphy.ac.cn; wangzi@bc.edu; hjgao@iphy.ac.cn.

Charge density waves (CDWs) are a widespread collective electronic order in quantum materials, furnishing key insights into symmetry breaking and competing phases. However, their dynamic control with external fields remains a pivotal challenge. Here, we report deterministic and hysteretic switching of unidirectional CDW orientation via in-plane magnetic field rotation in magnetic kagome metal $GdTi_3Bi_4$. Atomically resolved spectroscopy shows two types of $3a_0 \times 1a_0$ CDW domains, $Q_1$ and $Q_2$ oriented 60° apart along two distinct crystallographic directions and separated by atomically sharp domain walls. Rotating the magnetic field drives reversible transitions between these CDW configurations, exhibiting a robust $C_2$-symmetric phase diagram with pronounced hysteresis. This hysteretic switching is mediated by a field-dependent reorientation of underlying antiferromagnetic spins, revealing a tunable energy landscape with stable and metastable states and modulates the electronic charge order via spin-lattice coupling. Our findings not only demonstrate the switching of CDW configurations by in-plane magnetic field but also reveal the mechanism of coupling between CDW and magnetic fields, offering new insights into CDW manipulation and versatile platform for developing a spin-mediated multistate spin-charge coupling memory and programmable quantum devices.

Charge density waves (CDWs), characterized by periodic charge modulations in quantum materials, represent a fundamental type of symmetry-breaking order prevalent in correlated systems. They often coexist or compete with other emergent states such as superconductivity [1–5], electronic nematicity[6], and intertwined orders[7]. CDW phenomena have therefore been extensively investigated across a wide range of material families, including unconventional superconductors where CDW competes with superconductivity[8–14], kagome metals hosting unconventional topological charge orders[15,16], and low-dimensional materials[17,18]. External tuning parameters such as strain and magnetic fields have been reported to perturb the CDW configuration in certain compounds, for instance $NbSe_2$ [18-23]. Of particular interest are CDWs in magnetic materials[24-28], where the coupling between charge and spin degrees of freedom offers a rich platform for exploring new spin-charge coupling phenomena and controlling quantum states through magnetic fields[29–31]. In these systems, the CDW state is often either intrinsically pinned by the lattice or its manipulation results in a perturbative modification rather than a deterministic, nonvolatile switching of its order parameter. Thus, understanding this spin-charge coupling ultimately requires atomic-scale control and visualization of CDW dynamics in magnetic fields using systems where spin, charge, and lattice are intrinsically intertwined.

Recently, kagome magnets have emerged as model platforms for exploring correlated quantum phenomena, where their unique geometry hosts Chern and Weyl topological magnetism [32–35], anomalous Hall effect (AHE)[36–38], and CDWs arising from the interplay of multiple quantum orders. Geometric frustration and quantum interference lead to unconventional magnetic and charge orders, enabling direct coupling between spin and CDW phases [39–41]. A notable example is the $RTi_3Bi_4$ family (R = rare-earth metals), which exhibits frustrated one-dimensional (1D) zigzag spin chains decoupled from the kagome layers[42,43]. This unique architecture enhances magnetic tunability, resulting in novel correlated phenomena such as fractional magnetization plateau, spin density wave, and topological Hall effect (THE)[28,42–44]. Specifically, $GdTi_3Bi_4$ exhibits magnetization plateaus and, crucially, hosts field-tunable bi-oriented magnetic orders that can be precisely controlled via in-plane field rotation[45]. This makes it a promising platform for probing magnetically coupled electronic orders. However, it remains unclear whether CDWs emerge in this compound and, if so, whether their coupling to the magnetic orders enables deterministic and hysteretic switching of the CDW state via in-plane magnetic field rotation.

Here, we directly visualize the emergent unidirectional CDW in GdTi$_3$Bi$_4$ and demonstrate its deterministic control via in-plane magnetic fields using scanning tunneling microscopy/spectroscopy (STM/STS). We identify coexisting of $3a_0 \times 1a_0$ CDW domains oriented along two distinct Gd-Gd lattice directions. By rotating in-plane magnetic fields, we achieve controlled transitions between two orientations of CDW, with stripe orientations preferentially aligning with the applied field direction. Crucially, this switching process exhibits pronounced hysteresis, a signature of metastable states arising from the collective reorganization of underlying spin textures. The strongly spin-charge coupled electronic order establish conclusive evidence for spin-mediated mechanism and a tunable energy landscape with stable and metastable states, enabling nonvolatile multistate control of CDW configurations. Our findings provide a microscopic foundation for designing correlated spin-charge coupling memory devices and programmable quantum phases.

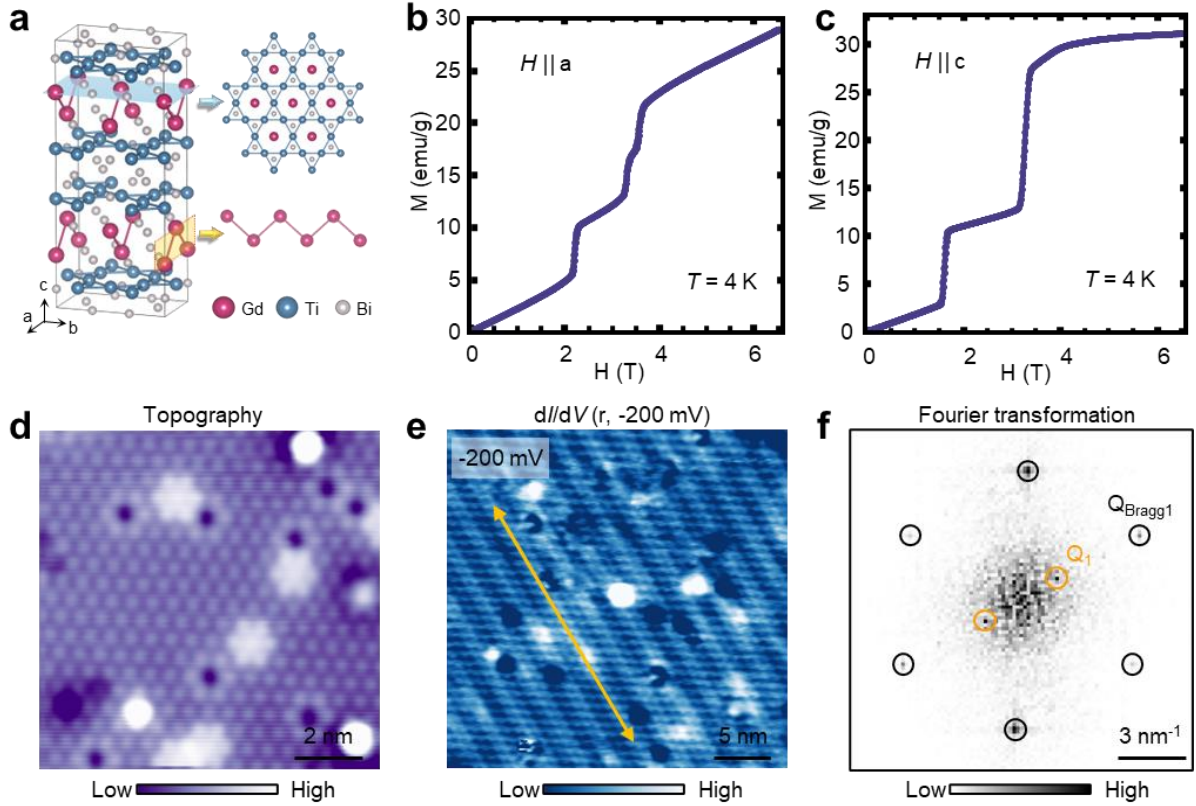

**Fig. 1 | Observation of canted spin texture and unidirectional CDW in GdTi₃Bi₄. a**, Crystal structure of GdTi₃Bi₄ showing the unit cell (left panel), Gd triangular lattice in the ab-plane (upper right panel), and Gd zigzag chain in the ac-plane (lower right panel). **b**, Magnetization measured at $T$ = 4 K with varying magnetic fields applied parallel to the a-axis. **c**, Magnetization measured at $T$ = 4 K with varying magnetic fields applied parallel to the c-axis. **d**, Atomically resolved STM topography of Gd-terminated surface (10 nm × 10 nm, $V_s$ = 1 V, $I_t$ = 100 pA). **e**, d$I$/d$V$ conductance map taken at 0 T revealing the unidirectional CDW. The orange arrow indicates the orientation of 3a-CDW (30 nm × 30 nm, $V_s$ = -200 mV, $I_t$ = 1 nA). **f**, A magnified view of the Fourier transform analysis of (e), showing the characteristic wavevectors $Q_1$ and $Q_{Bragg1}$.

Crystallographic analysis reveals that GdTi₃Bi₄ crystallizes in the *Fmmm* space group (No. 69), corresponding to a centrosymmetric orthorhombic lattice (Fig. 1a). Its three-dimensional architecture emerges from the synergistic coupling between two-dimensional Ti-based kagome layers and quasi-linear Gd chains arranged in a distinct zigzag pattern. Notably, the surface-projected Gd lattice exhibits a hexagonal lattice in top view, faithfully capturing the interlayer coordination environment surrounding the Gd ions. The magnetic sublattice is effectively decoupled from the kagome framework, thereby granting GdTi₃Bi₄ enhanced magnetic tunability.

Previous magnetization and transport measurements have established that GdTi$_3$Bi$_4$ exhibits an antiferromagnetic (AFM) ground state below a Néel temperature ($T_N$) of ~14.5 K[45]. The interplay between this AFM state and the quasi-one-dimensional Gd chain geometry generates significant magnetic frustration, leading to remarkably complex magnetic behavior. Specifically, there are fractional magnetization plateaus emerging along both a-axis and c-axis in GdTi$_3$Bi$_4$[42,43,46], as well-defined by magnetization measurements (Figs. 1b-c, and Fig. S1), serving as hallmark signatures of magnetic frustration. These findings strongly suggest the formation of a canted spin structure with multi-directional AFM orderings along different crystallographic axes, which may lead to unconventional physical properties.

Atomic-scale characterization through scanning tunneling microscopy (STM) provides further insights. Cleaved GdTi$_3$Bi$_4$ crystals expose Gd-terminated surfaces, as confirmed by high-resolution STM topography (Fig. 1d) showing well-ordered atomic periodicity with a lattice constant of 5.90 Å, which is well consistent with the crystallographic *a*-axis reported in previous studies[42]. Fourier analysis reveals sharp Bragg peaks, while d$I$/d$V$ linecuts show periodic intensity modulations (Fig. S2), demonstrating excellent sample crystallinity. Spatially resolved d$I$/d$V$ conductance map at 4.5 K (Fig. 1e) reveals a unidirectional charge density wave (CDW) with $3a_0 \times 1a_0$ configuration (indicated by orange arrow). Fourier analysis (Fig. 1f) identifies primary CDW (labeled Q$_1$) and Bragg (labeled Q$_{Bragg1}$) wavevectors, with Q$_1$ closely matching the expected 3a-CDW wavevector position. We use Q$_1$ to designate the nearly commensurate one-dimensional CDW configurations, where the charge stripes order is oriented -62° relative to the scanning x-direction.

Recent studies report four distinct CDW phases in GdTi$_3$Bi$_4$ under varying temperature and out-of-plane field[47], along with novel bi-oriented in-plane AFM domains[45]. Understanding the origin of the incommensurate unidirectional CDW phase and the role of complex in-plane AFM order represents a crucial challenge for unraveling CDW phenomena in strongly correlated systems.

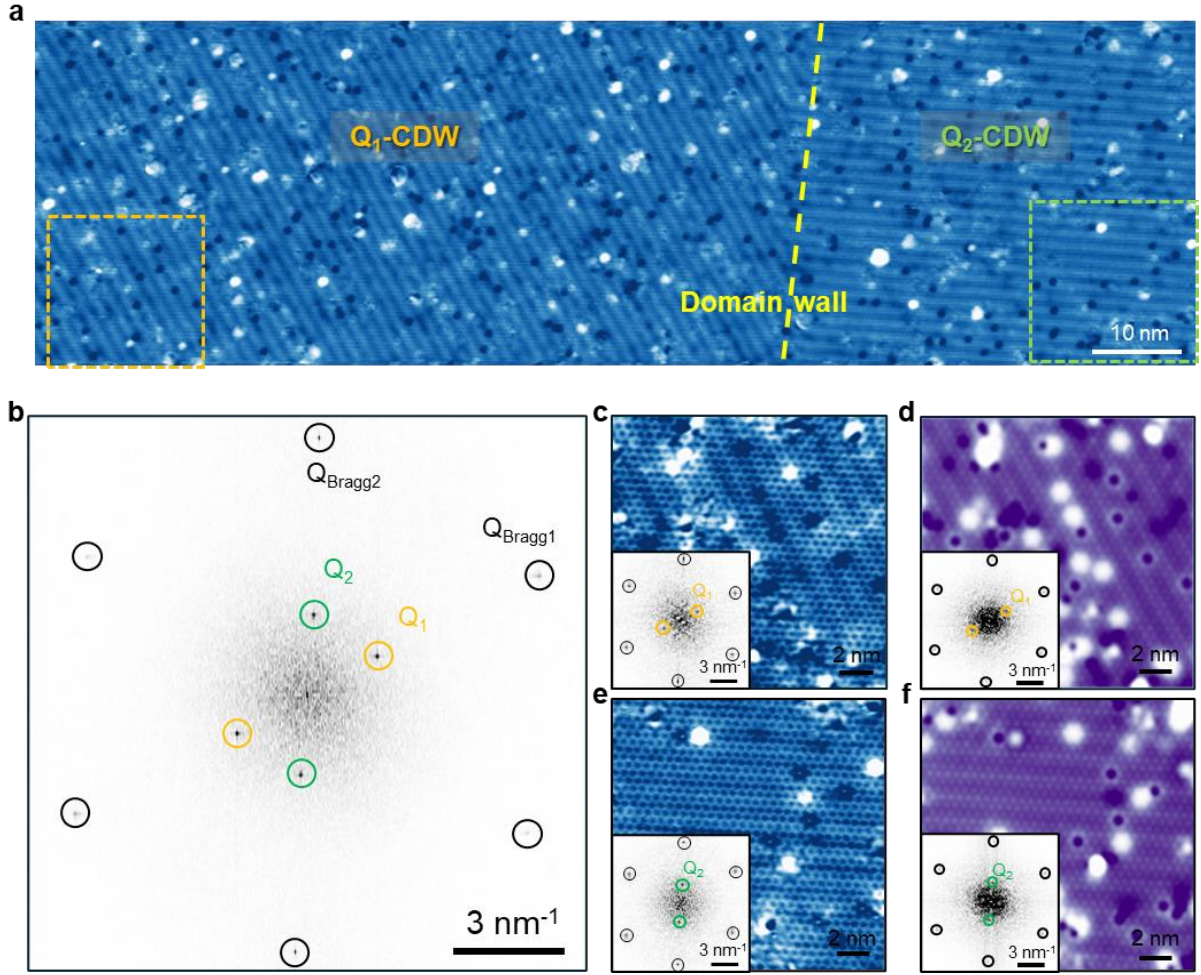

**Fig. 2 | Observation of domain wall with two CDW domains. a**, d$I$/d$V$ conductance map revealing two CDW domains with distinct orientations taken at 0 T (39 nm × 130 nm). The yellow dashed line indicates the domain wall. This image corresponds to a section of the area shown in Fig. S6a. **b**, Fourier transformation of Fig. 2a. The orange, green, and black circles marked the $Q_1$, $Q_2$, and $Q_{Bragg}$ peaks, respectively. **c-f**, d$I$/d$V$ conductance maps and atomically resolved STM topography taken at the same area of $Q_1$ domain (c, d) and $Q_2$ domain (e, f) taken at 0 T (17 nm × 17 nm). The regions are marked by orange (c, d) and green (e, f) boxes in (a). Insets: corresponding Fourier transformations.

The dramatic CDW behavior manifests in the coexistence of two types of 3a-CDW along two distinct directions: $Q_1$ at -62° (left region in Fig. 2a) and $Q_2$ at -2° (right region in Fig. 2a) relative to the scanning x-direction, corresponding to different principal Gd-Gd bond directions. The intervening domain wall (yellow dashed line) extends across the entire field of view while remaining decoupled from underlying topographic features (e.g., step edges). This domain wall morphology is different from the conventional density wave domain walls, which typically follow atomic-scale surface structure[24].

Fourier transform (FT) analysis (Fig. 2b) identifies two well-defined CDW modulations, $Q_1$ (orange circles) and $Q_2$ (green circles), exhibiting an about 60° angular offset. Corresponding real-space d$I$/d$V$ imaging and topography at atomic resolution simultaneously captures both the CDW patterns and underlying atomic structure within the designated orange and green regions in Fig. 2a, unambiguously establishing two spatially separated domains (Figs. 2c-2f). The CDW order is robust against the presence of point impurities. Measurements of $Q_1$ and $Q_2$ domains demonstrate that the CDW periodicities remain robust against impurities and vacancies. The sharp FT peaks from both domains confirm highly homogeneous, unidirectional CDW order. The CDW features remain robust and energy-independent (Figs. S3-S5). The coexistence of multi-directional CDW domains expands the understanding of unidirectional CDW phases[47], revealing that such unidirectional CDW can form along two degenerate Gd-Gd lattice directions, while implicitly excluding the zigzag chain direction along the a-axis.

Further measurements under out-of-plane magnetic fields demonstrate the tunability of CDW domains. At 3 T, both $Q_1$ and $Q_2$ signatures are fully suppressed, while reducing the field to 0 T restores only the $Q_1$ domain. This reconfiguration is accompanied by a domain wall displacement of approximately 70 nm and slight reconfigurations (Fig. S6). This sensitivity to the out-of-plane field arises from the gradual rotation of Gd magnetic moments towards the c-axis, which weakens the in-plane antiferromagnetic correlations essential for stabilizing the unidirectional CDW. Consequently, the external field serves as a direct knob to modulate both the out-of-plane spin component and spatial configuration of the charge order. Such a magnetically dominated response is fundamentally distinct from conventional CDW systems governed by lattice mismatches or charge correlations; it reveals an intrinsic spin-driven mechanism where the electronic charge order acts as a direct reporter of the underlying magnetic texture.

Crucially, the domain dynamics of the CDW exhibit a significant correlation with macroscopic magnetic behaviors[45]. The application of increasing out-of-plane magnetic fields was found to promote c-axis spin alignment while concomitantly weakening in-plane magnetic ordering. Notably, subsequent field reduction preserved the overall domain configuration while inducing localized rearrangements of domain walls. A pivotal discovery from the prior research is that the in-plane magnetic domains exhibit precise, field-tunable vectorial control under applied in-plane magnetic fields. This remarkable property implies that synergistic modulation of in-plane magnetic anisotropy may achieve unprecedented directional-selective switching of charge density wave states—a capability that could redefine control paradigms in correlated quantum materials.

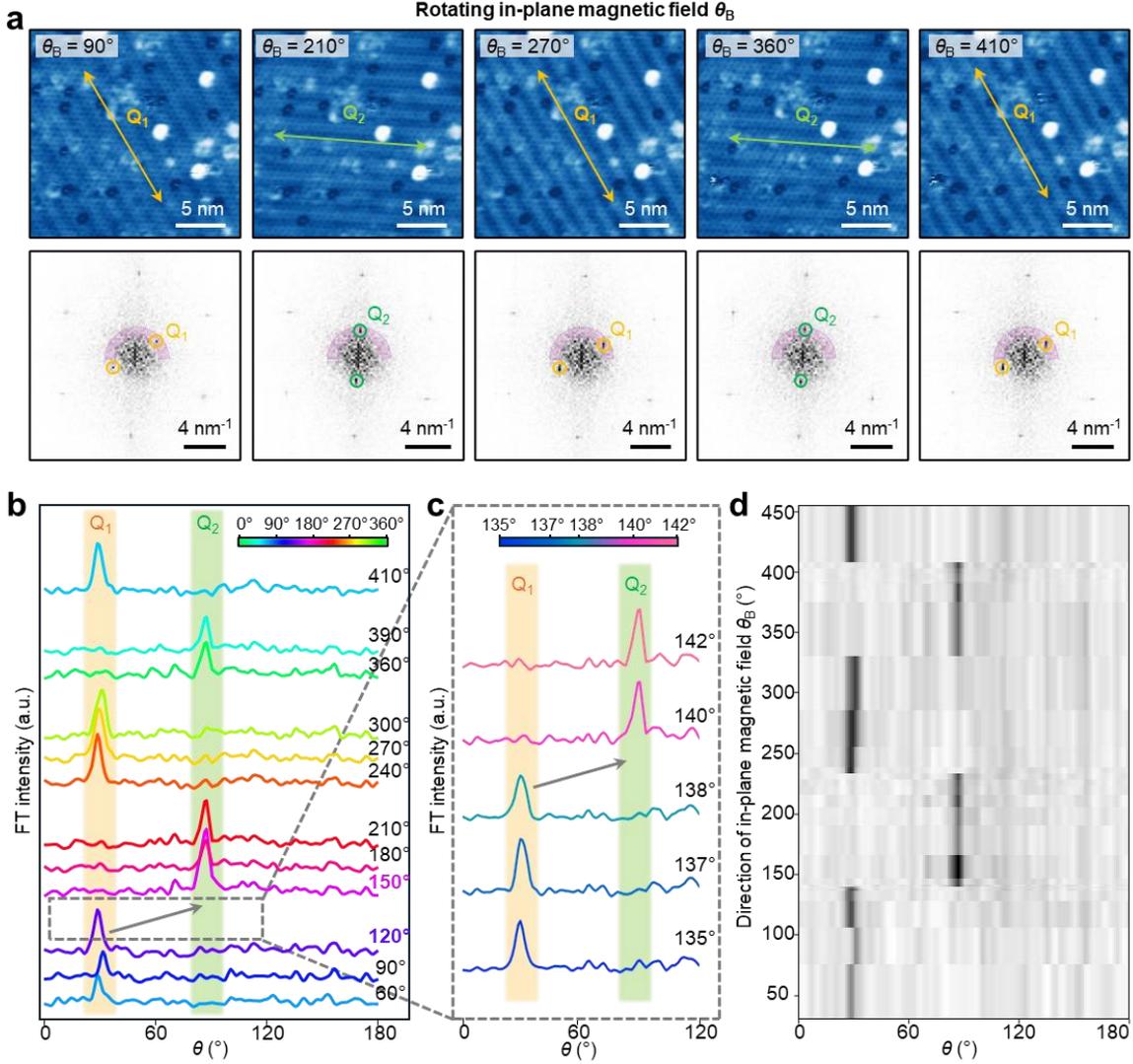

**Fig. 3 | Tuning and switching of Q$_1$ and Q$_2$ CDW configurations with in-plane magnetic field. a**, Upper panel: d$I$/d$V$ conductance maps under varying in-plane magnetic field, showing alternating Q$_1$ and Q$_2$ configurations (40 nm × 40 nm, $V_s$ = -0.2 V, $I_t$ = 1 nA). The CDW orientations are marked with orange and green arrows. Lower panel: the corresponding Fourier transformations. **b-d**, Angular dependence analysis of FT intensity profiles: waterfall representation (b), zoom-in waterfall representation around the switching point 120° - 150° (c) and color-code intensity map (d) generated by averaging the spectral weight along an arc-shaped sampling region (pink-shaded area). The x-axes represent the angular coordinate of sampling points.

To investigate the coupling mechanism between the CDW and the in-plane magnetic order, a series of STM and STS measurements are performed under a controlled, directional rotating in-plane magnetic field. Unlike the mixed-domain region shown in Fig. 2, a uniform and large-scale area (800 nm × 800 nm) exhibiting exclusively the Q$_1$ CDW configuration (Fig. S7) was selected. The prevalence of such single-domain regions in our experiments indicates a domain distribution spanning the micrometer-length scale. This targeted selection enabled us to precisely probe the magnetic field effects while eliminating artifacts from domain dynamics.

Focusing on a selected region, angle-resolved magnetic field measurements (30° rotational increments, 1.8 T purely in-plane field) reveal strong spin-charge coupling and a systematic evolution of CDW configurations (Fig. 3a). When the field is applied along the scanning y-direction ($\theta_B = 90°$), the $Q_1$ configuration remains completely undisturbed. In striking contrast, a distinct character presents at $\theta_B = 180°$, where the CDW wavevector switches from $Q_1$ to $Q_2$, demonstrating magnetic-field-driven reorientation as the field aligned closely with the $Q_2$ stripe orientation. Further rotation restores the $Q_1$ at 270°, followed by another switch to $Q_2$ at 360°. Ultimately, the system returns to its original $Q_1$ ground state with identical symmetry after a full rotation cycle. This reproducible switching between the two CDW wavevectors, exhibiting periodicity and $C_2$ symmetry, highlights the strong spin-charge coupling, wherein in-plane magnetic field modulates the CDW orientations. Detailed d$I$/d$V$ conductance maps (Fig. S8) further confirm the preferential alignment between the external field and stripe direction.

Quantitative analysis of the CDW order is achieved through the Fourier transform (FT) of differential conductance (d$I$/d$V$) maps under varying field orientations. The angular-resolved FT spectra, generated by averaging intensity along an arc-shaped sampling strip (radius matching the CDW wavevector magnitude, pink shaded region in Fig. 3a), clearly display the characteristic wavevectors of $Q_1$ (28°) and $Q_2$ (86°) peaks (Fig. 3b). Throughout the field rotation, the spectra consistently show single dominant peaks, indicating sharp transitions between the two wavevectors without intermediate-state coexistence. Specifically, high-resolution measurements near the switching points reveal that the CDW orientation flips abruptly within a single 2° step. This rapid reconfiguration, devoid of intermediate states or wavevector coexistence, provides clear evidence for a first-order-like transition in the electronic order (Fig. 3c and Fig. S9). The color-encode intensity map (Fig. 3d) categorizes the CDW wavevectors into five groups over 360°, revealing periodic oscillations in orientation. This systematic angular dependence, with unambiguous $C_2$ periodicity and high reproducibility, provides definitive evidence for the precise and reproducible control of CDW orientation via in-plane magnetic field direction. Moreover, these observations suggest that while the formation of unidirectional CDW breaks the inherent symmetry of GdTi$_3$Bi$_4$, the $Q_1$ and $Q_2$ states, once degeneracy-lifted, exist as stable and metastable configurations, whose relative energy difference can be tuned by the external field.

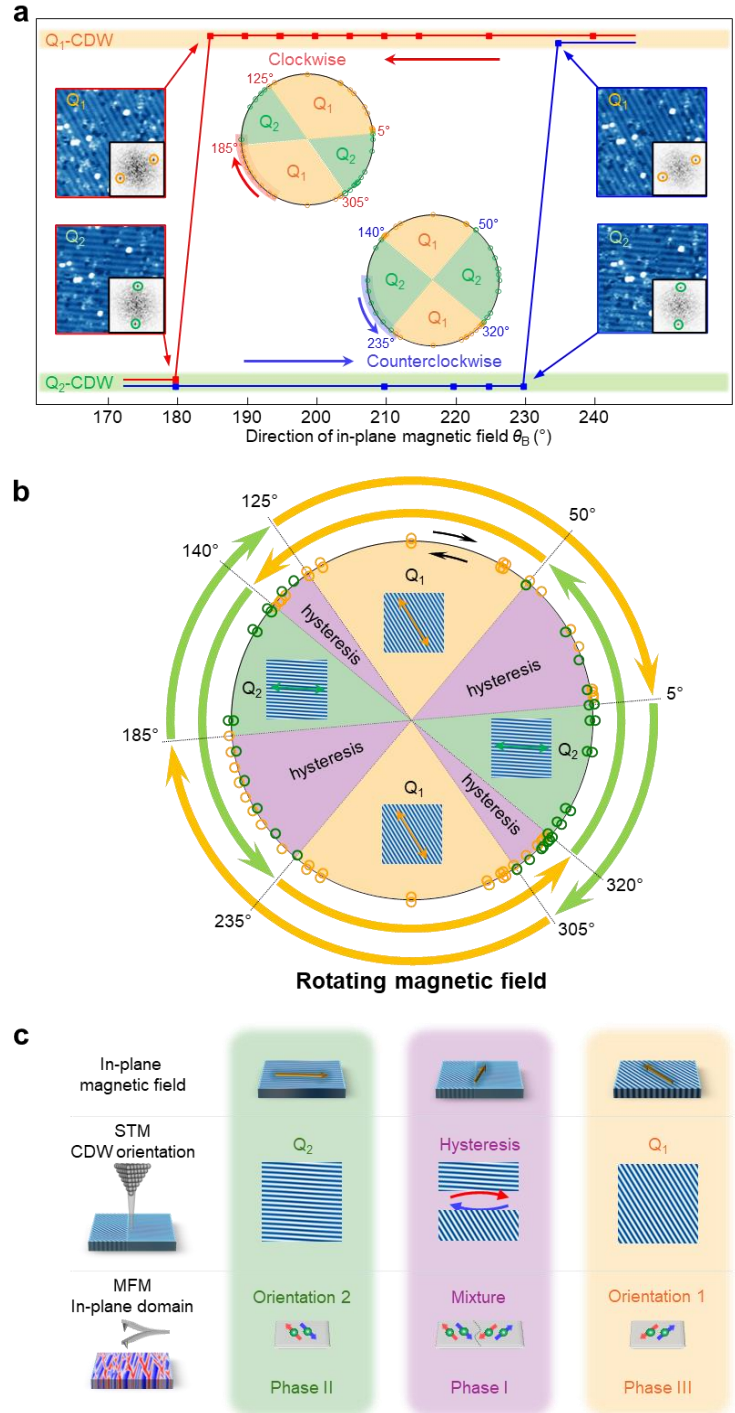

**Fig. 4 | Observation of hysteresis of CDW switching and phase diagram of CDW orientation. a**, Magnetic field rotation process showing distinct hysteretic behavior in orientation switching. The clockwise and counterclockwise processes are denoted by red and blue traces. Four representative d$I$/d$V$ conductance maps capture the configuration before and after switching transitions. Insets: Full angular rotation measurements for both rotational directions. **b**, Phase diagram of CDW configurations under a 1.8 T rotating in-plane magnetic field. The $Q_1$, $Q_2$, and hysteresis regimes are color-coded as orange, green and purple respectively. **c**, Comparison of the phases between CDW orientation states and in-plane magnetic domain configurations, revealing their cooperative ordering behavior.

Further investigation confirms that the field-driven CDW behavior is mediated by the reconfiguration of underlying spin textures and in-plane magnetization, rather than a directly response to the magnetic field. A crucial piece of evidence comes from the observed magnetic hysteresis associated with long-range magnetic order (magnetic domains). This fundamental distinction is strongly supported by the observed hysteresis effects under a rotating in-plane magnetic field, which are characteristic of long-range spin order dynamics and bistable energy landscape.

The most compelling evidence for this indirect coupling mechanism comes from our detailed examination of the wavevector transition process between 180° and 240°. We observe a pronounced hysteresis window of approximately 50° during continuous rotation of the 1.8 T in-plane magnetic field (Fig. 4a). This hysteretic behavior, which persists through multiple rotation cycles, provides unambiguous evidence for the existence of metastable states in the system's energy landscape and clearly demonstrates the crucial role of magnetic domain dynamics in mediating the field-CDW interaction. It is important to distinguish this angular hysteresis from conventional magnetic hysteresis. Standard *M-H* curves along fixed axes (Figs. 1b-c) show no hysteresis, as a characteristic feature of antiferromagnets with canted moments. In contrast, the hysteresis in Fig. 4 arises under a rotating-field protocol, where the direction of a fixed 1.8 T in-plane field is swept, which will be discussed below.

When the field is rotated in the counterclockwise direction (blue curves and squares in Fig. 4a, images in Fig. S10), the transition from $Q_2$ to $Q_1$ occurs at 235°. Conversely, during clockwise rotation (red curves and squares in Fig. 4a, images in Fig. S11), the system maintains its stable $Q_1$ state until reaching 185°, only undergoing the $Q_1$-to-$Q_2$ transition at 180°. The robustness and universality of this magnetic hysteresis phenomenon are compellingly demonstrated by the two polar plots (inset of Fig. 4a), which systematically document complete 360° rotation in both directions. This hysteresis between rotational pathways demonstrates a direction-dependent energy landscape featuring metastable states.

Through systematic analysis of multiple field rotation cycles in both clockwise and counterclockwise directions (Figs. S10-S13), we establish a comprehensive phase diagram of CDW states as a function of in-plane field orientation (Fig. 4b). While minor stochastic variations in the exact switching angles are observed across experimental repetitions, the diagram clearly identifies three distinct regimes modulated by the in-plane magnetic field. The majority of the angular range exhibits stable $Q_1$ or $Q_2$ configurations that remain unchanged with rotation direction, preserving $C_2$ rotational symmetry. A narrow intermediate regime exhibits strong rotation-direction dependence without a fixed CDW configuration. Notably, the angular extents of the $Q_1$ and $Q_2$ regions and their hysteresis windows are markedly asymmetric, reflecting underlying symmetry breaking. Although the three Gd-Gd crystallographic directions are nominally equivalent, the local crystal field and magnetocrystalline anisotropy break this symmetry, distinguishing the two a' directions, yielding unequal angular stability ranges for $Q_1$ and $Q_2$, in good agreement with our simulations (Fig. S14 and Supplementary Information Text II).

Critically, while the triangular arrangement of Gd atoms suggests three equivalent Gd-Gd directions, the interlayer Gd zigzag chain structure creates symmetry breaking, distinguishing the a-axis from the two a'-axis directions (±60°). We suggest that $Q_1$ and $Q_2$ configurations align with the a'-axis orientations, while the a-axis direction cannot support charge stripe order due to its distinct intra-chain magnetic configuration. This structural symmetry breaking accounts for both the absence of a third CDW state and the dominance of $Q_1/Q_2$ configurations in the phase diagram.

Our atomic-scale CDW observations show remarkable alignment with mesoscopic magnetic domain behavior, with both techniques identifying three phases under in-plane field rotation[45]. Rotating the in-plane field establishes a direct one-to-one correspondence between the STM and MFM phase diagrams (Fig. 4c). The $Q_1$ and $Q_2$ CDW regions, with reproducible configurations across measurements, correspond to single-domain in-plane antiferromagnetic (AFM) orders in phases II and III, respectively , demonstrating that the CDW switching is governed by the dynamics of the underlying antiferromagnetic spin textures. Conversely, the hysteretic CDW regime reflects the spatial randomness of magnetic domains, matching the mixed AFM domain configuration in phase I, directly implicating a tunable energy landscape in which the relative stability of bistable states is modulated by the field orientation.

The above findings establish a definitive microscopic picture of a spin-mediated unidirectional CDW in GdTi$_3$Bi$_4$, characterized by robust, long-range, and energy-independent LDOS modulations, which is the definitive spectroscopic fingerprints of charge ordering. The unidirectional CDW stripes exclusively form along two degenerate crystallographic directions (Q$_1$ and Q$_2$), energetically forbidden along the Gd-Gd zigzag chain direction (a-axis) due to its distinct magnetic environment. This selective formation reflects a magnetically induced C$_2$-symmetric ground state. The deterministic and hysteretic switching of the CDW orientation between the Q$_1$ and Q$_2$ states, achieved via in-plane magnetic field rotation, constitutes a central finding of this study. Besides orientation control, the evolution from 1D to 2D CDW under out-of-plane fields (Fig. S6) enables continuous tuning of the charge order's dimensionality. The contrasting responses of the CDW to out-of-plane and in-plane magnetic fields reveal two distinct control mechanisms. Out-of-plane fields primarily modulate the spatial texture of the CDW by suppressing in-plane AFM correlations, whereas in-plane rotating fields directly control the CDW orientation by selecting between degenerate AFM easy axes. While both effects originate from spin-charge coupling, they operate through fundamentally different physical pathways.

Crucially, the hysteretic switching between the Q$_1$ and Q$_2$ CDW configurations is not a direct field effect but is mediated by the reorientation of the underlying in-plane antiferromagnetic spin order. This reorientation of the Néel vector, driven by the rotating in-plane magnetic field, establishes a tunable energy landscape with stable and metastable states, where their relative stability is governed by the field orientation relative to the magnetic easy axes. The locking between the CDW wavevector and this spin configuration is naturally achieved via spin-lattice coupling. Both the antiferromagnetic order of the Gd moments and the 3a$_0$ CDW induce periodic lattice distortions, making the lattice the essential mediator between spin and charge. To minimize the total free energy within this landscape, the CDW wavevector realigns itself as the field-modified spin texture alters the symmetry of the spin-induced lattice distortion. This microscopic picture is directly supported by the one-to-one correspondence between the angular phase diagram of CDW orientations from STM and the AFM domain phase diagram measured by MFM, which share nearly identical switching angles and hysteresis widths (Fig. S14).

The robust coupling between the CDW orientation and the field-dependent spin texture, facilitated by spin-lattice interactions, gives rise to a novel spin-charge intertwined density wave that is cooperatively responsive to external magnetic fields. This spin-charge coupling memory effect enables the nonvolatile, deterministic switching of a global unidirectional CDW order parameter. By rotating the in-plane magnetic field, we induce hysteretic switching between the $Q_1$ and $Q_2$ configurations. This hysteresis reveals a tunable energy landscape featuring stable and metastable CDW states, with their relative stability dictated by the field orientation relative to the AFM easy axes. The strong coupling between the CDW Q-vector orientation and the field-dependent spin reorientation establishes a robust magnetoelectronic memory effect (supported by theoretical modeling in Fig. S14 and Supplementary Text II). These field-induced, bistable CDW configurations provide a direct pathway towards novel device concepts. The hysteretic switching of the Q-vector orientation, encoding distinct states, forms the foundation for realizing multistate nonvolatile spin-charge coupling memory elements in spintronics and quantum information technologies.

While our work reveals a strong coupling between magnetic order and CDW modulation and its control via in-plane magnetic fields, elucidating the precise spin structure in $GdTi_3Bi_4$ requires further investigation. Although spin-lattice coupling offers a plausible explanation, an alternative mechanism involving a primary $6a_0$ spin density wave (SDW) driving the $3a_0$ CDW merits consideration. We propose neutron scattering, which has recently been applied to $CeTi_3Bi_4$[44], or other experimental techniques to further confirm the spin structure and understand the evolution of complex magnetic orders[45], including the periodically of the spin structures, the in-plane spin direction (±7° orientations), and the 1/3 magnetization plateaus along the a- or c-axis directions.

We show the hysteretic and nonvolatile control of $3a_0 \times 1a_0$ CDWs in frustrated kagome magnet $GdTi_3Bi_4$. The atomically resolved measurements demonstrate that the orientation of CDW domains is directly governed by the field-dependent reorientation of the underlying antiferromagnetic spin texture. The hysteretic transitions between two crystallographic directions under a rotating magnetic field, and the resulting $C_2$-symmetric phase diagram, provide direct evidence of a tunable bistable energy landscape born from spin-charge interplay. The unambiguous one-to-one correspondence between CDW configurations and magnetic domain dynamics provides conclusive evidence for a spin-charge intertwined density wave. This spin-mediated spin-charge coupling switching represents a paradigm for the control of correlated electronic phases. Our findings establish a new mechanism for CDW manipulation and spin-charge coupling memory effect, paving the way for the development of functional devices that leverage spin-charge coupling for quantum information technology.

# Methods

**Single-crystal growth of GdTi$_3$Bi$_4$ sample**

GdTi$_3$Bi$_4$ single crystals were grown using a self-flux method. Gd powder (Alfa 99.9%), Ti powder (Alfa 99.99%), and Bi (Alfa 99.997%) were measured in a 2:3:12 ratio into an alumina crucible. The crucible was sealed in an evacuated silica ampoule. The sealed ampoule was heated in a furnace to 1000 °C over 12 h, held at the temperature for 24 h, and slowly cooled down to 500 °C at a rate of 2 °C/h, where excess bismuth was removed through centrifugation. Hexagonal shape, shiny-silver single crystals with a size up to 3×3×1 mm$^3$ could be easily harvested at the bottom of the alumina crucible. The samples deteriorate quickly in humid air, and the surface of the sample will tarnish even in an argon-filled glovebox.

**STM/STS characterization**

The samples used in the experiments are cleaved at room temperature (300 K) and immediately transferred to an STM chamber. Experiments were performed in an ultrahigh-vacuum (1 × 10$^{−10}$ mbar) ultralow-temperature STM system equipped with a 9-2-2 T magnetic field. All of the scanning parameters (setpoint voltage and current) of the STM topographic images are listed in the figure captions. Unless otherwise noted, the d$I$/d$V$ spectra were acquired by a standard lock-in amplifier at a modulation frequency of 973.1 Hz. Tungsten tips were fabricated via electrochemical etching and calibrated on a clean Au(111) surface prepared by repeated cycles of sputtering with argon ions and annealing at 500 °C.

**Transport Measurement**

Electrical transport measurements were collected on a Quantum Design Physical Properties Measurement System (PPMS). The measurements of magnetic property were performed on a Quantum Design magnetic property measurement system (MPMS-3).

# Acknowledgments

The work is supported by grants from the National Natural Science Foundation of China (Nos. 62488201, 12522408, 12374199, 92580202, 12522407, 12174424), the National Key Research and Development Projects of China (Nos. 2022YFA1204100, 2025YFA1212800), and the

Beijing Nova Program (Nos. 20240484651). Z. W. acknowledges the support of the U.S. Department of Energy (DOE), Basic Energy Sciences (BES), Grant No. DE-FG02-99ER45747 and Cottrell SEED Award No. 27856 from the Research Corporation for Science Advancement.

## Author contributions

H.-J.G. supervised and coordinated the project. S.Z. and H.-J.G. designed the experiments. Z.C., S.Z., and K.X. conducted STM measurements and analyzed data. R.W. and H.Y. synthesized the single crystals. Z.W. provides theoretical support. N.W., J.S., and J.C. conducted magnetization measurements. Z.C., S.Z., and J.G. wrote the manuscript with inputs from all authors.